# Structure and morphology evolution of concave-shaped SiC {0001} surfaces in liquid silicon

Xinming Xing[1,a], Takeshi Yoshikawa[2,b] and Didier Chaussende [1,c *]

[1]Univ. Grenoble Alpes, CNRS, Grenoble INP (Institute of Engineering), SIMAP, 38000 Grenoble, France

[2]Institute of Industrial Science, The University of Tokyo, 4-6-1 Komaba, Meguro-ku, Tokyo 153-8505, Japan

[a]xinming.xing@grenoble-inp.fr, [b]t-yoshi@iis.u-tokyo.ac.jp, [c]didier.chaussende@grenoble-inp.fr

**Abstract.** Concave-shaped 4H-SiC {0001} surfaces have been prepared and reconstructed in pure liquid silicon for investigating the structure and morphology evolution of the SiC surface as a function of both azimuthal and off-axis angles. Different surface characteristics are revealed on two polar surfaces where only the Si face reflects the six-fold symmetry of 4H-SiC crystal. On the Si face, the step bunching along the $<1\bar{1}00>$ direction is stronger than the $<11\bar{2}0>$ direction, which is related to the bonding state at the step edge. More significant step bunching is observed on the C face whereas it is not sensitive to azimuthal orientation. The extent of step faceting is stronger on the Si face. The step faceting is independent of the off angle on both polarities of SiC {0001} surfaces.

## Introduction

Vicinal surfaces refer to the crystal surfaces being cut at a relatively small angle to one of the low index surfaces, where the step spacing and step configuration are determined by the off angle and off-axis orientation. Vicinal SiC surfaces have been utilized in vapor phase epitaxy, which is the key to replicate the polytype through the so-called "step controlled epitaxy" [1]. However, as the density of steps is high, the steps are usually unstable due to their interactions with neighbors and with environment (fluid). As a consequence, large macrosteps of several micrometers in height are frequently observed in the solution growth of 4H-SiC, which then cause the occurrence of trench defects and solvent inclusions [2]. Therefore, it is of great importance to understand the step dynamics on vicinal surfaces. For example, it has been reported that step and surface morphological instability was closely related to the off angle and crystalline orientation of the seed crystal [3][4].

To get a full picture of the effect of crystalline orientation on surface reconstruction, a series of experiments are normally required. This is usually both material- and time-consuming. Besides, it is not easy to prepare ideal smooth SiC substrates with specific crystalline orientations. Therefore, we implemented an original approach, which consisted in using a concave-shaped SiC surface. It allows investigating a large variety of vicinal surfaces in one single experiment.

In this work, the concave-shaped SiC surfaces of two polarities are prepared and reconstructed during dissolution in liquid silicon at high temperature; the surface morphology is characterized and its evolution features are presented.

## Experimental

The concave-shaped surfaces were prepared by using a ball crater grinder (Calotest CAT$^2$c, Anton Paar). The instrument is designed for measuring the thickness of single or multi-layer coatings, whose principle is schematically illustrated in Fig. 1. Generally a small crater is formed on the fixed SiC surface after being ground with a stainless steel ball of known geometry (*R*). In our case, only one outline can be detected on the surface after the grinding, whose diameter is marked as *r*. Both the Si- and C-faces of 4H-SiC nominal {0001} wafers were used as the substrates. The actual off angle of the substrates was 0.2° from the AFM characterization. The used polishing ball was 30 mm in diameter, rotating at a rate of 900 rpm. Two different diamond slurries of 0.5-1 μm and 0-0.2 μm were used sequentially for obtaining a relatively smooth wall of the crater. The final diameter of the crater was about 2.3 mm, which gave the off angle relative to the nominal surface ($\theta^0$) as equal to 4.3°. Considering the original off angle, the obtained concave-shaped surfaces owned the in-plane azimuthal angle ($\varphi$) varying from 0° to 360° and the out-of-plane polar angle ($\theta$) from -4.5° to 4.1°.

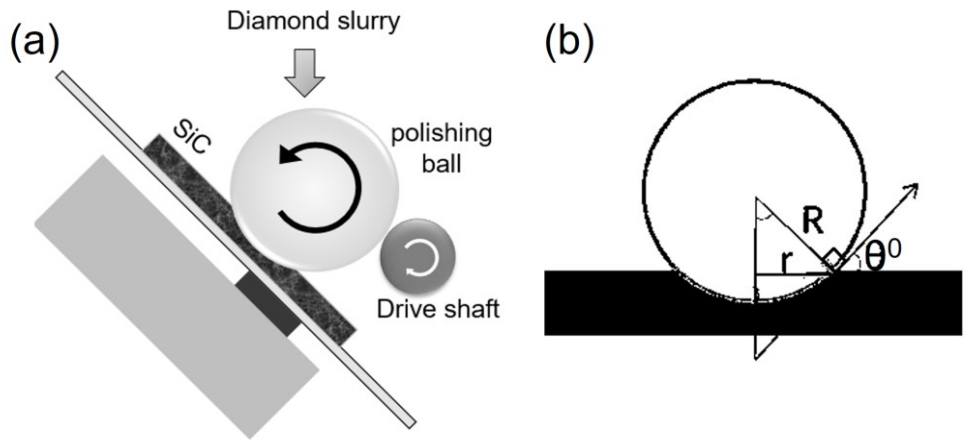

Figure 1 Schematic illustration of the surface preparing process by using the crater grinder. The off angle can be calculated from the relationship $sin(\theta^0) = r/R$ where *r* is the diameter of the crater seen from the top and *R* is the diameter of the grinding ball.

The prepared substrates were washed in acetone and in deionized water for 10 min separately, by using an ultrasonic cleaner. A small electronic-grade silicon piece was put on the crater area at the beginning and then the SiC/Si liquid interface was formed after melting of silicon at high temperature. The sample was processed in argon gas (900 mbar) at 1700 °C for 30 min. At the end of the surface reconstruction, the liquid silicon was quickly removed by a tool before natural cooling down. This made the surface morphology generated at high temperature to be preserved for further ex-situ characterization. The reconstructed surface was thoroughly characterized by DIC optical microscope and AFM.

**Results and Discussion**

The surface morphologies of two polar concave-shaped surfaces after being reconstructed at 1700 °C for 30 min were scanned under DIC microscopy and are presented in Fig. 2. For both cases, the concave surfaces have been reconstructed into clear step-and-terrace structures, and the step spacing and shape are changing with both the radial and azimuthal directions. On the Si face, the morphology and size distribution of steps reflect the six-fold symmetry of the SiC crystalline structure. However, the surface structure is more homogenous on the C face, exhibiting a perfectly concentric morphology. The two representative azimuthal directions, <1$\bar{1}$00> and <11$\bar{2}$0>, can be distinguished according to the shape of the central part, whose six edges are perpendicular to the <1$\bar{1}$00> direction (only on the Si face). The steps moving along <1$\bar{1}$00> direction are straighter and sparser. While along the <11$\bar{2}$0> direction, or in the transition areas between two <1$\bar{1}$00> areas, the steps are waving and crossing. A similar trend was reported around the screw dislocation centre when the 6H-SiC (0001) Si face surfaces were etched by $H_2$ [5]. However, it should be noticed that the steps formed on our surfaces are much larger than the unit cell level steps formed upon $H_2$ etching.

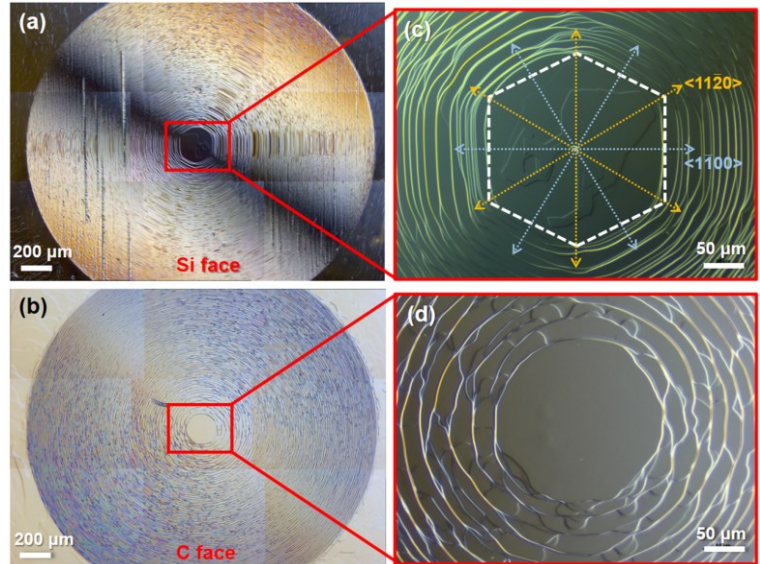

Figure 2 DIC pictures of two concave-shaped 4H-SiC surfaces reconstructed at 1700 °C for 30 min. (a) the Si face and (b) the C face. (c) (d) the corresponding enlarged central areas.

The detailed step structures were characterized by AFM and are shown in Fig. 3. Along with both the <1$\bar{1}$00> and <11$\bar{2}$0> directions of the Si face, an obvious increase of the step density is observed with the increasing off angles. The tendency is found a bit more significant along the <1$\bar{1}$00> direction. Here the waving of steps is not clear even along the <11$\bar{2}$0> direction because the AFM scan area is too small to show the long range step shape. For the C face, it is difficult to distinguish two crystalline orientations thus only one random orientation is selected. The step density on the C face is much smaller than that on the Si face for all the off angles investigated. It should be noticed that the step density only slightly increases with the increasing off angles on the C face. Combined with the DIC pictures, we primarily conclude that the surface structures on the C face are only weakly dependent on the surface orientation including azimuthal angle and off angle.

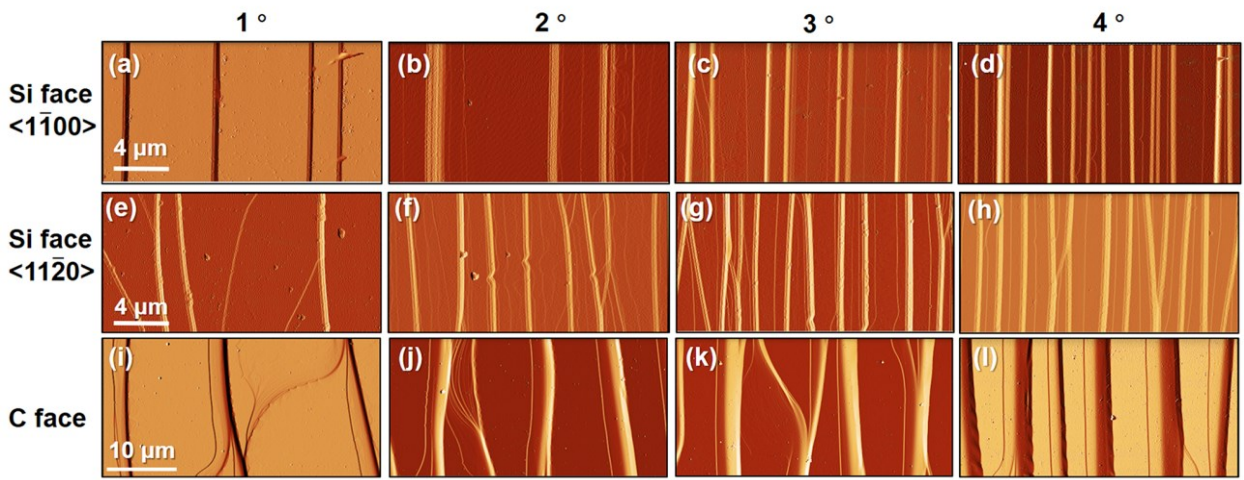

Figure 3 AFM amplitude images showing the step-and-terrace structures on the concave-shaped surfaces. (a)~(h): 4H-SiC Si face and (i)~(l): 4H-SiC C face. The horizontal scanning lengths are 20 μm and 40 μm for the Si face and the C face, respectively.

Furthermore, the step and terrace structures are quantitatively described based on the AFM results. The step density, the average step height and the macrostep angle were extracted from the height profiles and are shown in Fig. 4. The step density is the reciprocal of the step spacing while the macrostep angle is defined as the acute angle between the macrostep edge and lower terrace (as illustrated in Fig. 4c). As we can see, the height of the macrosteps is seemingly independent of the off angle on the Si face. The C face shows a gradual increase of the step height when increasing the off angle. Combining with the almost constant step density, it is supposed that the step bunching on the C face is much stronger than that on the Si face. However, the macrostep angles on the C face are always smaller than the Si face and almost invariant with off angle. This indicates the step faceting on the C face is much weaker, which is not dependent on the crystalline orientation as well.

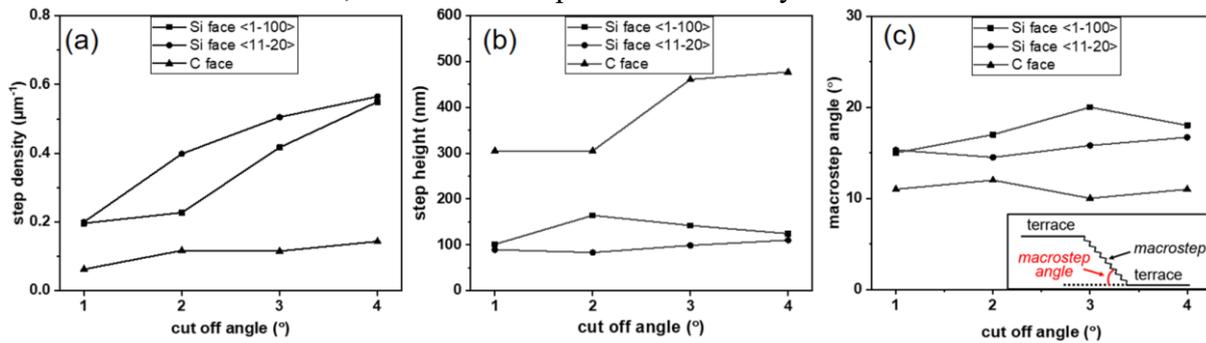

Figure 4 Dependences of surface step parameters on the cut off angle along different crystalline orientations. (a) step density, (b) step height, and (c) macrostep angle. The inset of Fig. 4c skematically shows the defination of the macrostep angle.

On the Si face, the extent of bunching seems constant everywhere along both $<1\bar{1}00>$ and $<11\bar{2}0>$ orientations. Yet, it should be noticed that the step bunching along the $<1\bar{1}00>$ of the Si face is always more pronounced than the $<11\bar{2}0>$ direction. This might be related to the bonding state at the step edge. For the steps arranged perpendicular to the $<1\bar{1}00>$ direction, every two bilayer steps alter their characters along the c-axis between one dangling bond and two dangling bonds. The steps with two dangling bonds should be receding faster and then the step bunching results. When the direction deviates from $<1\bar{1}00>$ direction, the total dangling bonds number at the step edge varies with the changing azimuthal angle. While reaching the $<11\bar{2}0>$ direction, all the steps have the same dangling bond state and thus propagate at a homogeneous rate leading to a weaker step bunching.

**Summary**

We presented the application of concave-shaped SiC surfaces in studying the effect of crystalline orientation on surface reconstruction in liquid silicon. The reconstructed Si face revealed a six-fold symmetric structure which was not observed on the C face. The extent of step bunching was more pronounced on the C face and enhanced with off angle while it was almost invariant on the Si face. The step bunching was gradually degrading from the $<1\bar{1}00>$ direction to $<11\bar{2}0>$ direction on the Si face. The step edge termination model was used to interpret the evolution of the step bunching along the azimuthal direction. The extent of step faceting is more significant on the Si face.